\documentclass[prd,twocolumn,showpacs,nofootinbib,amsmath,amssymb,floatfix,superscriptaddress,showkeys]{revtex4}
\usepackage{multirow}
\usepackage{graphicx}
\usepackage{epstopdf}
\usepackage{dcolumn}
\usepackage{bm}
\usepackage{threeparttable}
\usepackage{subfigure}
\usepackage{color,txfonts}
\usepackage{ulem}
\definecolor{blue0}{rgb}{0,0,0.6}
\usepackage[colorlinks,linkcolor=blue0,anchorcolor=blue0,citecolor=blue0,urlcolor=blue0]{hyperref}

\newcommand{\jcap}{J. Cosmol. Astropart. Phys.}

\newcommand{\mnras}{Mon. Not. R. Astron. Soc.}

\newcommand{\aap}{Astron. Astrophys}
\newcommand{\aj}{Astron. J.}

\newcommand{\apjl}{Astrophys. J.}

\newcommand{\beq}{\begin{equation}}
\newcommand{\eeq}{\end{equation}}
\newcommand{\beqa}{\begin{eqnarray}}
\newcommand{\eeqa}{\end{eqnarray}}

\begin{document}
\title{Revisiting the Geminga halo at GeV energies with Fermi-LAT data}
	
\author{Yu Meng}
\affiliation{Guangxi Key Laboratory for Relativistic Astrophysics, School of Physics Science and Technology, Guangxi University, Nanning 530004, China}
\author{Yun-Feng Liang}
\email[]{ liangyf@gxu.edu.cn}
\affiliation{Guangxi Key Laboratory for Relativistic Astrophysics, School of Physics Science and Technology, Guangxi University, Nanning 530004, China}
\author{Ben-Yang Zhu}
\affiliation{Guangxi Key Laboratory for Relativistic Astrophysics, School of Physics Science and Technology, Guangxi University, Nanning 530004, China}
\author{Xiao-Na Sun}
\affiliation{Guangxi Key Laboratory for Relativistic Astrophysics, School of Physics Science and Technology, Guangxi University, Nanning 530004, China}
\author{En-Wei Liang}
\email[]{lew@gxu.edu.cn}
\affiliation{Guangxi Key Laboratory for Relativistic Astrophysics, School of Physics Science and Technology, Guangxi University, Nanning 530004, China}
	
\begin{abstract}
Nearby pulsars within $\sim1\,{\rm kpc}$ are considered to be possible sources of 10-500 GeV cosmic-ray positron excess measured by PAMELA and AMS-02. A TeV halo around Geminga is detected by HAWC, and the measurements of its surface brightness profile indicate a slow particle diffusion surrounding the source. This result challenges the pulsar interpretation of the positron excess. The observations at GeV energies provide direct information on the electron/positron density in the GeV nebula, which can offer more direct constraints on the origin of the positron excess. 
Two previous works have performed analyses on the GeV emission of the pulsar halo, but focused on the energy band above 8 GeV. 
In this work, we use a longer dataset from the Fermi Large Area Telescope (LAT) to re-analyze the GeV halo emission of Geminga, extending the analysis to cover the energy range of 1-1000 GeV. 
We find that the analysis in this wider energy range results in a low significance of the halo emission. 
This can be attributed to the Galactic interstellar emission model being unable to perfectly fit the background over this broader energy range, and due to the low measured halo flux at $<$ 10 GeV energies leading to a mismatch between the observation and model expectation.
We also derive the spectral energy distribution of the tentative halo emission, which shows a very hard spectrum in the 1-10 GeV range.
\end{abstract}
	
\maketitle
	
\section{introduction}
Cosmic-ray positrons are considered to be related to the secondary product given by spallation reactions of primary cosmic rays (CRs) with atoms in the interstellar medium (ISM).
An excess of positron fraction is detected by PAMELA \citep{pamela09positron}, Fermi-LAT \citep{lat12positron} and AMS-02 \citep{ams13positron} in the energy range of tens to hundreds of GeV. The rapid energy loss of these sub-TeV CR electrons implies that the excess positrons should be produced by nearby sources. The explanation for the origin of the positron excess is still under discussed. One of the possible origins is the injection from nearby pulsars \citep{2009JCAP...01..025H,2009PhRvL.103e1101Y,2012CEJPh..10....1P,2013PhRvD..88b3001Y,2017JCAP...09..029J}. Among this scenario, Geminga is a very promising candidate for the positron excess due to its appropriate distance (250 pc) and age ($\sim3.5\times10^5$ years) \citep{ATNFpaper,Caraveo96_distpm}.

The Milagro gamma-ray observatory detected an extended $\gamma$-ray emission of $\sim2.6^\circ$ surrounding Geminga from 1 to 100 TeV \citep{milagro09}. The existence of this TeV gamma-ray halo is confirmed by the observation of the High-Altitude Water Cherenkov Observatory (HAWC) in the 5-40 TeV energy range, with an extension of $\sim2^\circ$ \citep{hawc17,HAWC:2024scl}. The halo is also detected by the H.E.S.S. telescopes \citep{HESS:2023sbf}. These measurements indicate a large amount of high-energy $e^+/e^-$ is released from the pulsar, and the gamma-ray emission of the halo is attributed to the inverse Compton scattering (ICS) of these $e^+/e^-$ with ambient photons of the interstellar radiation field (ISRF).
TeV gamma-ray halos generated with the same mechanisms are also detected around the sources Monogem by HAWC \citep{hawc17} and PSR J0622+3749 by the Large High-Altitude Air Shower Observatory (LHAASO) \citep{lhaaso21j0622}. 
There is also evidence that the radio-quiet pulsar PSR J0359+5414 is surrounded by a TeV halo \citep{HAWC:2023jsq}.
TeV halos around millisecond pulsars have also been studied \citep{hooper18msp,hooper21msp,HAWC:2025xjs}.

The HAWC observation also provides information on the angular surface brightness profile (SBP) of the TeV halo around Geminga, which suggests a lower diffusion coefficient ($\sim3\times10^{27}\,{\rm cm^2\,s^{-1}}$ at $100\,{\rm TeV}$ \cite{hawc17}) within at least a few tens of parsecs around the pulsar compared to that in the normal ISM ($\sim10^{30}\,{\rm cm^2\,s^{-1}}$ at 100 TeV \cite{Moskalenko:1997gh}). This result places important constraints on the models of accounting for positron excess with nearby pulsars \citep{2018ApJ...863...30F,2018PhRvD..97l3008P,dimauro19,xi19}, especially in the scenario of one diffusion zone model. However, the $e^+/e^-$ population that generates the TeV halo is not directly related to that contributing to the positron excess in the 10-500 GeV energy range. Instead, the latter is expected to produce halo emission in the GeV band. In view of this, two previous works \citep{dimauro19,xi19} have carried out analyses to search for the GeV halo emission around the Geminga pulsar with Fermi-LAT data. \citet{xi19} (X19 hereafter) derived upper limits on the GeV halo emission for various model parameters for both one- and two-zone models, while \citet{dimauro19} (D19 hereafter) claimed a detection of an extended emission around Geminga in the energy range above 8 GeV through analyzing a larger region around the source and taking into account the proper motion of the pulsar. Both works conclude that the Geminga pulsar can contribute only a few percent of the positron excess in the one-zone diffusion model.

We note that the two previous works only analyzed the Fermi-LAT observations of $>$8 GeV or $>$10 GeV. However, in fact, Fermi-LAT has a good overall performance in the 1-10 GeV range. In this work, we use a longer data set (13 years) from Fermi-LAT to re-analyze the GeV emission of the Geminga halo, and extend the adopted data to the range of 1-1000 GeV. As we will see below, the 1-10 GeV data is in fact crucial for identifying models. We also attempt to use only the data corresponding to an off-pulse period of Geminga to reduce the influence of the pulsar's contamination in the $<10$ GeV range. 
Our analysis aims to explore whether the detection of the GeV halo of Geminga is robust with this larger data set, and constrain the model parameters (e.g., index of the electron spectrum, diffusion coefficient) through the obtained halo spectrum in 1-1000 GeV.

\section{Fermi-LAT analysis}
\label{sec:analysis}
Fermi-LAT is a large field of view ($\sim2.4$ sr), pair conversion telescope, sensitive to gamma-ray photons between $\sim$30 MeV and $>$500 GeV \citep{atwood09lat}. This work uses 13 years of Fermi-LAT data from 2008-08-04 to 2021-07-14  (MET 239557417-647991919)\footnote{\url{https://fermi.gsfc.nasa.gov/cgi-bin/ssc/LAT/LATDataQuery.cgi}}. We adopt the data belonging to the Pass 8 SOURCE event class and use the corresponding P8R3\_SOURCE\_V2 instrument response functions (IRFs). We employ the {\tt Fermitools}\footnote{\url{https://fermi.gsfc.nasa.gov/ssc/data/analysis/software/}} to perform a binned analysis on Fermi-LAT data. We select photons with energies from 1 GeV to 1000 GeV within a $50^\circ$ region around the position of Geminga at (RA, Dec)=(98.48, 17.77). Applying a zenith angle cut of $z_{\rm max}<90^\circ$ allows us to effectively exclude the contamination of the photons from the Earth's limb\footnote{We have tested that alternatively using a $z_{\rm max}$ of 105$^\circ$ has a very minor influence on our main results.}. We also use {\tt gtmktime} tool to select good time intervals defined by the expression {\tt (DATA\_QUAL>0 \&\& LAT\_CONFIG==1)}. The data are binned into counts cube with the {\tt gtbin} tool, with the pixel size and energy binning listed in Table \ref{tab:likers}.

\begin{table*}[!t]
\centering
\caption{\label{tab:likers} Information on likelihood analysis of the Geminga halo.}
\begin{tabular}{ccccccc}
\hline
\hline
{Parameters}   & {Config. I} & {Config. II}  & {Config. III} & {Config. IV}   & {Config.~V} & {Config.~VI}   \\
               & (All data)  & (All data)    & (Off-pulse)  & (Di Mauro+2019) & (Xi+2019)    & (HPX binning) \\
\hline
{ROI$^{a}$}                & $70^\circ\times70^\circ$ & -$^{b}$             & $70^\circ\times70^\circ$ & $70^\circ\times70^\circ$ & $50^\circ\times50^\circ$ & $40^\circ$ radius \\
{Pixel size}               & 0.06$^\circ$             & -             & 0.06$^\circ$             & 0.06$^\circ$             & 0.05$^\circ$             & NSide=512$^{c}$   \\
{$N_{\rm bins,dec}$$^{d}$} & 6                        & -             & 6                        & 6                        & 6                        & 6                 \\
{$E_{\rm min}$ [GeV]}      & 1                        & -             & 1                        & 8                        & 10                       & 1                 \\
{$E_{\rm max}$ [GeV]}      & 1000                     & -             & 1000                     & 1000                     & 1000                     & 1000              \\
{$T_{\rm stop}$ [MET]}     & 647991919                & -             & 647991919                & 541871021                & 541871021                & 647991919         \\
{$R_{\rm src}$$^{e}$}      & 53$^\circ$               & -             & 53$^\circ$               & 53$^\circ$               & 43$^\circ$               & 43$^\circ$        \\
{$Z_{\rm max}$}            & 90$^\circ$               & -             & 90$^\circ$               & 90$^\circ$$^f$               & 105$^\circ$              & 90$^\circ$        \\
SP$_{\rm IEM}$$^{g}$             & Const                    & Powerlaw      & Const                    & Const                    & Const                    & Const             \\
\hline
\multicolumn{7}{c}{log-Likelihood}\\
\hline
{proper}    & -2995426.9 & -2995164.4 &-3211960.1 & -473029.8 & -216172.8  & -1820538.3 \\
{no proper} & -2995426.2 & -2995162.4 &-3211960.1 & -473037.0 & -216176.7  & -1820538.3 \\
{no halo}   & -2995427.0 & -2995175.4 &-3211960.1 & -473061.4 & -216181.4  & -1820538.3 \\
\hline
\multicolumn{7}{c}{TS}\\
\hline
{proper}    & 0.2 & 21.9 & 0 & 63.2 & 17.2 & 0 \\
{no proper} & 1.6 & 25.9 & 0 & 48.8 & 9.4  & 0 \\
\hline
\end{tabular}
\begin{tablenotes}
\item $^{a}$ {Region of interest. For WCS-based analysis (i.e., I-V), we use CAR projection in the CEL coordinate system.}
\item $^{b}$ The same as the left column. 
\item $^{c}$ {NSide=512 corresponds to an $\sim0.11^\circ$ mean spacing of the pixels.} 
\item $^{d}$ {Number of bins per energy decade.}
\item $^{e}$ {Radius within which 4FGL-DR2 sources are included in the model file.}
\item $^{f}$ In Ref.~\cite{dimauro19}, they in fact used $z_{\rm max}<105^\circ$. We have tested that alternatively using a $z_{\rm max}$ of 105$^\circ$ has a minor influence on the results.
\item $^{g}$ The scaling spectrum of the standard IEM, where {\tt Const} represents a {\tt ConstantValue} spectrum (i.e., no spectral scaling).
\end{tablenotes}
\end{table*}

We include in the background model the Galactic diffuse emission (IEM) and isotropic diffuse emission (ISO), as well as all sources listed in the LAT 10-year Source Catalog (4FGL-DR2) \citep{4fgl,4fgldr2} within a $R_{\rm src}$ region (see Table \ref{tab:likers}). The two diffuse components are modeled with the templates {\tt gll\_iem\_v07.fits} and {\tt iso\_P8R3\_SOURCE\_V3\_v1.txt}.
We first perform a background-only fit without the halo component included in the model ({\it no halo}). The package {\tt Fermipy}\footnote{\url{https://fermipy.readthedocs.io/en/latest/}} \citep{fermipy} is used to implement the global binned likelihood analysis. 
Considering that the ROI contains too many sources, especially for the $70^\circ\times70^\circ$ ROI\footnote{There are in total 682 sources and 1248 free parameters within a 40$^\circ$ radius centered on Geminga.}. Allowing all parameters of all sources to vary simultaneously would lead to a very-long-time computation and poor convergence. Therefore, we alternatively employ the following strategy to perform the fitting. We first optimize all source parameters (including both normalization/prefactor parameters and spectral shape parameters) in the model source by source with the {\tt optimize()} function in Fermipy. 
This process involves a complete loop over all sources in the ROI (including point, extended, and diffuse sources), and one complete loop is referred to as one ROI optimization. The ROI optimization procedure is repeated several times (typically 10-15 times under our convergence criterion) until the likelihood value becomes steady (i.e., the change in log-likelihood between two optimizations is less than 1, $\Delta\ln\mathcal{L}<1$).
After this {\tt optimize} step, the parameters of all sources are already very close to optimal. Based on this optimized model, we then perform two rounds of global likelihood fit. In the likelihood fit, only normalization parameters of the two diffuse sources (i.e., IEM and ISO) and prefactors of 4FGL sources with ${\rm TS}>100$ within 20$^\circ$ circle are free to vary. For all the cases in our analysis, we find the {\tt optimize} step already gives good enough results, and the likelihood value does not increase much in the {\tt fit} step. This strategy is necessary to avoid issues arising from an excessive number of free parameters in the {\tt fit} step.

Based on the best-fit model from the background-only fit, we add the halo component to the model.
We generate {\tt mapcube} templates of the halo emission for the Fermi-LAT data analysis according to the calculation of electron propagation and ICS radiation (see Appendix \ref{sec:cal} for the calculation details).
Since for the energies considered in our analysis, the slow-diffusion zone dominates the ROI and different halo models would generate similar spatial morphologies of the gamma-ray ICS halo \citep{dimauro19}, for simplicity, we assume a one-zone diffusion model for the halo. 
We construct the templates adopting the optimal halo parameters in \cite{dimauro19} (i.e., $\gamma_e=-1.8$ and $D_0=2.3\times10^{26}\,{\rm cm^2/s}$, where $\gamma_e$ is the index of the injection electron spectrum and $D_0$ is the diffusion coefficient at 1 GeV), which are determined through fitting the surface brightness profile of the TeV halo and the 8-1000 GeV Fermi-LAT data.
A complete set of benchmark parameters for the halo ICS emission is listed in Table~\ref{tab:pars}.
The template is in a {\tt mapcube} format to include both spatial and spectral information of the halo model flux.
With the halo component included in the model, we perform the same {\tt optimize+fit} procedure as described above. For the halo, we free only the normalization parameter\footnote{Note that the halo parameters listed in Table~\ref{tab:pars} are used to construct the halo template and are not free parameters in the Fermi-LAT data analysis.}. The test statistic (TS) of the halo is defined as ${\rm TS}=-2(\ln L_{0}-\ln L)$, where $L_{0}$ is the maximum value of likelihood for the background-only model and $L$ is the maximum value of likelihood for the model with halo component added.

The proper motion of Geminga has a large impact on the morphology of the GeV halo emission.
We perform the analysis for both considering and not considering the pulsar's proper motion. 
In the case with proper motion included, the source position $\vec{r}_s$ is varied with time. The proper motion for Geminga is 178.2$\pm$1.8 mas/year \citep{ATNFpaper,Caraveo96_distpm}.
A {\it proper} template is for considering Geminga's proper motion, while a {\it no proper} one is for not considering.

\section{Results of likelihood analysis}
\subsection{Global likelihood fit}
We first attempt to reproduce the results of D19 \cite{dimauro19} and X19 \cite{xi19} (see cases IV and V in Table \ref{tab:likers}). The purpose of this step is to check the reliability of our analysis process and to see if there are any changes to the results by using the new source catalog and the new IEM template of Fermi-LAT.
{We use 4FGL-DR2 catalog and {\tt gll\_iem\_v07.fits} template file rather than 4FGL and {\tt gll\_iem\_v06.fits} in the two previous works.}
We find that the halo TS values we get are in good agreement with these two previous works, indicating that our data analysis is reliable.
Our results also confirm that there are signs of halo emission around the Geminga pulsar at $\gtrsim10$ GeV energies. 
The TS values for the {\it proper} model are generally larger than the {\it no proper} one. Since the Geminga pulsar actually has a proper motion with a velocity of 178.2$\pm$1.8 mas/year, the {\it proper} template aligns better with the physical reality. 
Therefore, detecting a higher TS value for the {\it proper} model than for the {\it no proper} model would support the possibility that the detected extended emission around Geminga is real from the ICS halo.
We note that the main difference between cases IV and V is the use of different ROI sizes, indicating that due to the large spatial extension of the halo emission, the selection of ROI will affect the results.

For this reason, our results of the 1-1000 GeV analysis are based on an ROI of $70^\circ\times70^\circ$.
For the 1-1000 GeV analysis (case I in Table \ref{tab:likers}), we find that very low TS values (${\rm TS}\sim0$) are obtained for both the {\it proper} and {\it no proper} models (but please see also the discussion below regarding the Config.~II case which gives higher TS values), although we have used observation data with a longer time period (13 years) and a wider energy range (begin with 1 GeV).
However, if we changed the energy range of our data from 1-1000 GeV to 10-1000 GeV (not changing the time period), a larger TS value similar to the results of cases IV and V is obtained.

\subsection{Some tests on the global fitting results}

In the energy range below a few tens of GeV, the Geminga pulsar emits very strong gamma-ray emission, which will raise the background for the analysis of the Geminga halo and may influence our analysis results.
To reduce the pulsar's flux, we also perform an analysis that uses only the data corresponding to the off-pulse period of Geminga to reduce the background (case III in Table \ref{tab:likers}). 
We define $\phi\in[0.03,0.18]\cup[0.55,0.65]$ as the on-pulse period and the rest as the off-pulse period (see Appendix~\ref{sec:phase} for an illustration of the phase-folded light curve), where $\phi$ is the pulse phase. We find that the likelihood analysis of the data in the off-pulse phase interval yields results consistent with those based on all data, and the TS values are $\sim$0 for both {\it proper} and {\it no proper} models, suggesting that the low TS value is not due to the pulsar's contamination. We have also tested that masking 2$^\circ$ regions around the 3 brightest sources (i.e., Geminga, Crab, and IC 443) has no influence on the obtained results of the halo.

In the above analysis, both the IEM and ISO diffuse components used the default model files and configurations from the Fermi collaboration\footnote{\url{https://fermi.gsfc.nasa.gov/ssc/data/access/lat/BackgroundModels.html}}. When analyzing large-scale extended emissions like the Geminga halo, the modeling of these diffuse components might affect the results. The diffuse emission at low Galactic latitudes is mainly dominated by the IEM component. Here we test the impact of the IEM model.
The standard Galactic interstellar emission model provided by the Fermi-LAT collaboration contains patches that absorb residuals mainly on the Galactic plane. These patches may also absorb part of the gamma-ray halo. 
Therefore, we test performing the analysis by replacing the standard IEM with a dust template or 8 alternative IEM templates. 
Dust can trace the distribution of interstellar gas, so the dust map can effectively serve as a template for the gas-correlated gamma-ray emission.
We construct the dust template from the Planck dust opacity map \cite{2011A&A...536A..19P}. 
The 8 alternative IEM templates are provided by the 1st Fermi-LAT Supernova Remnant Catalog\footnote{\url{https://fermi.gsfc.nasa.gov/ssc/data/access/lat/1st_SNR_catalog/}} \cite{Fermi-LAT:2015xeq}.
These tests are based on the Config.~I setup. We find that the analyses with these non-standard templates generally fit the Fermi-LAT data worse, i.e., the resulting likelihood values are typically smaller than those obtained using the standard IEM template. 
For example, for the dust template, we get a maximal $\ln(L)$ of $-3044455.0$, which is smaller than the value of $-2995426.2$ listed in Table I. The TS values obtained are 0.6 for the {\it no proper} case and 0 for the {\it proper} case, indicating that no significant halo component can be detected when using the dust map as the interstellar emission model.
For the 8 alternative templates, similar results can be obtained, i.e., no halo signal is detected in these tests (note the tests are based on Config.~I setup). Together with the impact of the IEM models on the SED results discussed in Sec.~\ref{sec:sed}, we can conclude that the choice of interstellar emission models does not introduce an artificial halo signal or mask a true excess.

In the above analyses, we use the default {\tt xml} model configuration of the IEM component, which frees only the normalization of the IEM template and fixes the spectral slope. We also test the effect of allowing the spectral slope of the Galactic interstellar emission model to vary (also based on the Config.~I analysis). The TS values obtained become 25.9 for the {\it no proper} case and 21.9 for the {\it proper} case, as can be found in the Config.~II column of Table~\ref{tab:likers}. This increase in TS further highlights the importance of properly modeling the interstellar emission for the detection of the GeV ICS halo of Geminga. 

Choosing different halo model parameters (e.g., $\gamma_0$ and $D_0$) will change the predicted gamma-ray spectrum of the halo emission. However, creating model templates based on these parameters requires complex calculations, so we cannot directly fit parameters such as $\gamma_0$ and $D_0$ in the Fermi-LAT data analysis. To allow more flexibility for the spectrum in the halo template file so that it can be better compatible with the observation, we attempt to include an additional log-parabola function to scale the gamma-ray spectrum in the halo template, allowing it to have more variation from its benchmark shape.
We find that after including the log-parabola scale, the fitting results show no further improvement compared to the Config.~II case\footnote{The fits using the log-parabola model often suffer from serious convergence issues, but we tested a range of initial parameters for the log-parabola and did not obtain higher likelihood values.}. 
We can not obtain TS values of the halo component larger than the ones reported in Table~\ref{tab:likers}.
In Sec.~\ref{sec:discuss}, we will further investigate the spectrum of the halo emission..

\begin{figure}[t]
\centering
\includegraphics[width=0.45\textwidth]{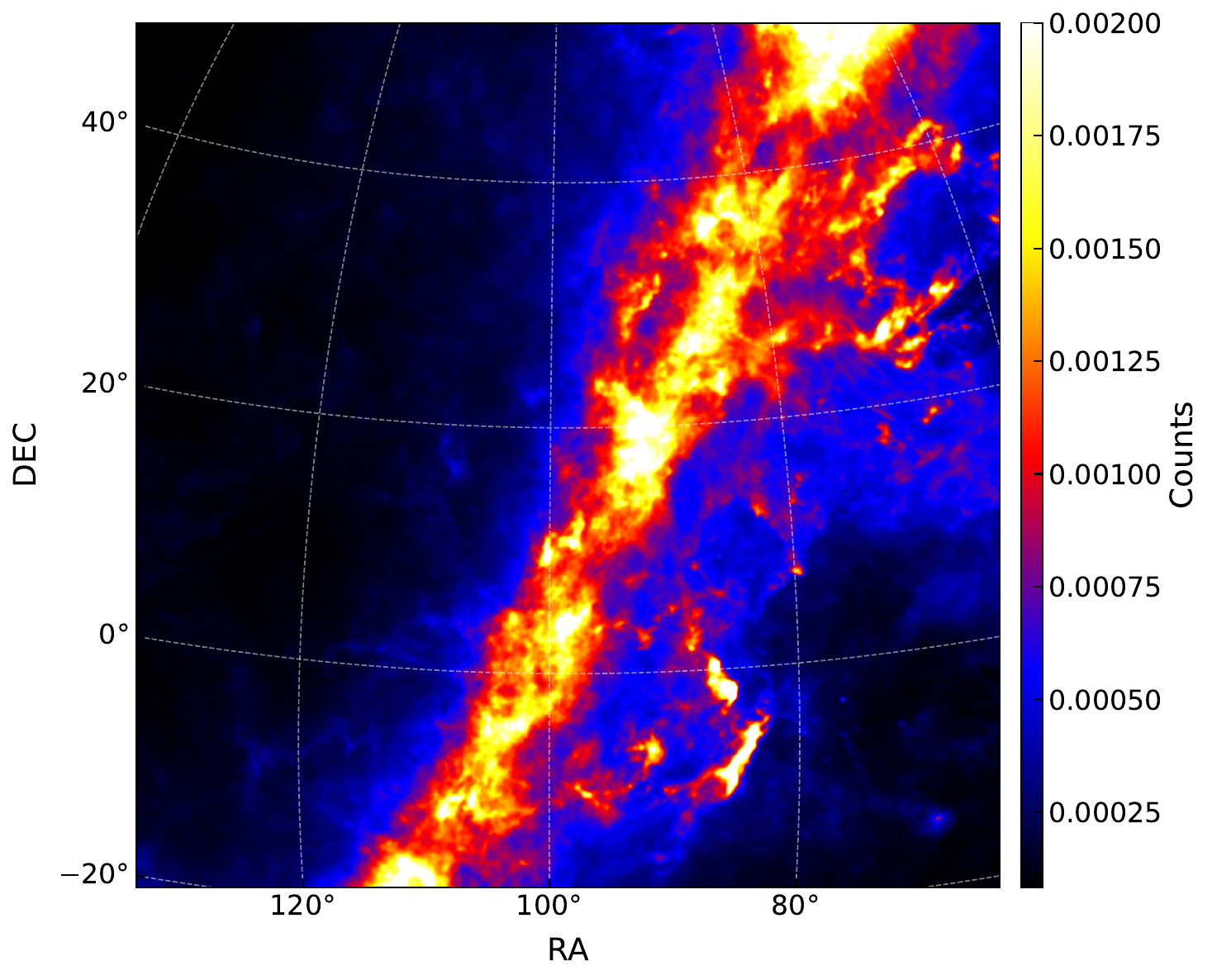}
\caption{Difference between the model maps derived from the best-fit IEMs of Config.~II and Config.~IV (M2$-$M4). The values are all positive, indicating that in Config.~II, the IEM background above 10 GeV is slightly elevated to accommodate the newly included $<10$ GeV data, causing lower TS values of halo in Config.~II than IV.}
\hfill
\label{fig:mmdiff}
\end{figure}

\begin{figure*}
\centering
\includegraphics[width=0.6\textwidth]{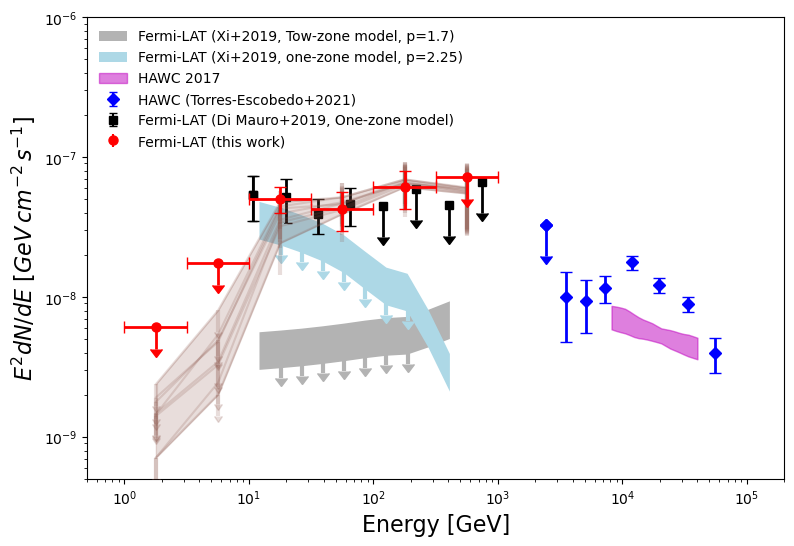}
\caption{SEDs for the Geminga ICS halo. Our results in the 1-1000 GeV energy range are shown as red points. Upper limits are reported when TS$<4$. 
The brown points are derived using 8 alternative IEM templates.
Also shown are the results from \cite{hawc17}, \cite{dimauro19}, \cite{xi19} and \cite{TorresEscobedo:2021xz}.}
\label{fig:sed}
\end{figure*}

We speculate that another reason for the decrease in the TS value of the halo component is that the background in the $>10$ GeV energy range has been elevated in the 1-1000 GeV analyses. To verify this, we further examine the best-fit IEM models from Config.~II and from Config.~IV. We generate model maps in the 10-200 GeV range\footnote{The tentative halo emission mainly appears in 10-200 energy range (see Fig.~\ref{fig:sed}), therefore the results are mostly affected by the background in this range.} for the two best-fit IEM models (denoted as M2 for Config.~II and M4 for Config.~IV). Fig.~\ref{fig:mmdiff} shows the difference between the two model maps (M2 - M4). It can be seen that the difference is positive, indicating that, in the 10-200 GeV range, the best fit IEM model of Config.~II is higher than that of Config.~IV. This is because adding the 1-10 GeV data slightly lifts up the high-energy part of the IEM model (a lever-like effect). Since the excess emission associated with the halo is quite weak, we believe the slightly increased background in the 10-200 GeV range suppresses the halo component, leading to a lower TS value.
Combining the results of Fig.~\ref{fig:mmdiff} and Config.~II, we conclude that the IEM is the most important factor responsible for the decrease in the TS value.

The above results (Config.~I-V) are derived using binned sky maps of WCS-CAR projection \citep{2002A&A...395.1061G}. However, for the data analysis of a large sky region, using WCS-based maps may cause large distortions for spatial pixels far from the ROI center. We therefore alternatively perform the likelihood analysis using the HEALPix (HPX) \citep{2005ApJ...622..759G} binning scheme (Config.~VI in Table~\ref{tab:likers}). We find that the obtained TS values are consistent with those of Config.~I. We have also checked that an 8-1000 GeV analysis with the HPX binning gives results consistent with Config.~IV. To avoid possible bias\footnote{Another benefit of using HPX binning is that in this case, our analysis ROI becomes a circular region. A circular ROI with a 40-degree radius covers about the same sky area as a square ROI of $70^\circ\times70^\circ$. However, we only need to add 4FGL-DR2 sources within $R_{\rm src}=43^\circ$ in the model file for analysis, compared to $R_{\rm src}=53^\circ$ required for the $70^\circ\times70^\circ$ ROI. This greatly reduces computational costs, especially the memory usage when running {\tt gtsrcmaps}.}, {\it in the following analysis, we will always use the HEALPix binning scheme.}

\subsection{Spectral energy distribution}
\label{sec:sed}
We next derive the spectral energy distribution (SED) of the pulsar halo. The SED is generated by performing the likelihood analysis bin-by-bin to derive the halo fluxes and corresponding uncertainties in different energy bins. We derive the SED based on the best-fit model of Config.~VI. In each energy bin, we free only the normalization/prefactor parameters of the halo, Crab, IC 443, Geminga pulsar, and the two diffuse components, and fix all other background sources. The SED is shown in Figure~\ref{fig:sed}, together with the spectra reported in the previous literature. 
As can be seen in the plot, above 10 GeV, the fluxes we derive are consistent with those reported by D19 \cite{dimauro19}. However, at energies below 10 GeV, it is shown that the halo has a very low flux. The $E^2{\rm d}N/{\rm d}E$ spectrum increases rapidly from $<10^{-8}\,{\rm GeV\,cm^{-2}s^{-1}}$ to $\sim10^{-7}\,{\rm GeV\,cm^{-2}s^{-1}}$ in the range of 1 to 10 GeV.
It seems that the halo ICS flux is suppressed at these lower energies and there is a cutoff in the spectrum. 

Since the standard Galactic IEM template includes patches that may absorb the Geminga halo emission and affect the SED results, we test using the 8 alternative templates from the 1st Fermi-LAT SNR catalog \cite{Fermi-LAT:2015xeq} to derive the halo SED.
These alternative templates contain HI and CO gas components. For the HI and CO, we fix the HI1, HI2, CO1, and CO2 components during the likelihood fitting process because they primarily belong to the inner Galactocentric rings and are not dominant in our region of interest. Freeing them in the fitting would cause convergence issues.
The SEDs derived using the 8 alternative templates are demonstrated as brown points in Fig.~\ref{fig:sed}. As shown, the results are in good agreement with the SED derived from the standard IEM template, also showing low fluxes below 10 GeV. Therefore, the SED results are robust against the test of using different IEM templates.

\begin{figure*}
\centering
\includegraphics[width=0.95\textwidth]{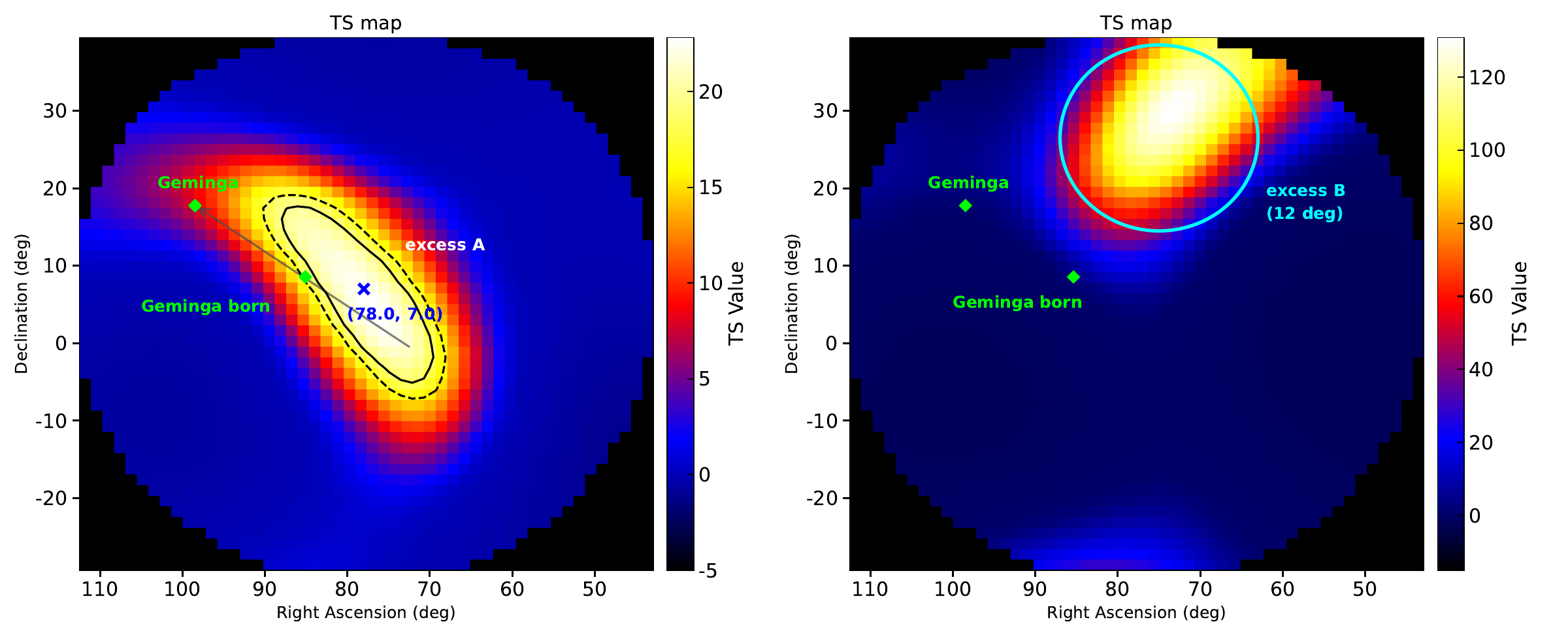}
\caption{The 10-1000~GeV ({\it left}) and 1-10~GeV ({\it right}) TS maps used to examine the existence of a halo-like component. {\it left:} The test source is a two-dimensional spatial Gaussian with $\sigma=10^\circ$ (see the main text and Appendix~\ref{sec:ext} for details on the determination of $\sigma$). The vector demonstrates the direction of the proper motion of the Geminga pulsar. The solid and dashed contours are for 68\% and 95\% confidence levels, respectively. {\it Right:} The test source is a 2D Gaussian with $\sigma=12^\circ$. The blue circle shows the extension of the best-fit Gaussian of the excess B.
These two TS maps cover a sky region within a 35-degree radius. They are generated using the Fermi-LAT data from a circular ROI with a 40-degree radius (i.e., following the Config.~VI analysis setup). 
Note that these are not standard point-source TS maps as derived using {\tt gttsmap}, but instead TS maps that show test results for an additional extended source component centered at each pixel of the map.
}
\hfill
\label{fig:tsmap1}
\end{figure*}

\section{Discussion}
\label{sec:discuss}
Although the above analysis shows possible $>10$ GeV emission from the halo, we still need to examine whether the tentative halo emission is at exactly the right place, or is at a position that is offset from the model expectation. If the latter is true, it is possible that the detected signal is a result of imperfect modeling of the Galactic diffuse background in the neighbouring region rather than a real halo signal. Therefore, we create a TS map to examine the existence of a halo-like component at each location of the ROI. For each spatial pixel of the TS map (we adopt a HEALPix binning), we put a putative radial Gaussian component into the model and derive the TS value of the Gaussian component to check if a halo-like component is present at that location. Here we use the radial Gaussian model to represent a potentially existing halo component, because the spatial morphology of the halo is approximately Gaussian-like. Our goal is to examine whether there is any extended excess component within this region, and if so, to determine its central position and spatial extent, in order to assess whether it is consistent with a Geminga halo or whether it may contaminate the analysis of the Geminga halo if it is unrelated to Geminga.
The extension of the Gaussian is chosen to be $\sigma = 10^\circ$, which corresponds to the $\sigma$ value that yields maximal TS value in the extension analysis (please see Appendix \ref{sec:ext} for the determination of the $\sigma$ of the Gaussian). As can be seen in the left panel of Figure~\ref{fig:tsmap1}, a tentative excess emission (denoted as ``excess A") does exist close to the Geminga born position. The fact that the excess is mainly present in the vicinity of the Geminga born position (rather than the Geminga position) can explain why the analysis gives lower TS values when using a smaller ROI or when not considering the proper motion (i.e., Config.~V or X19).

\begin{figure}[b]
\centering
\includegraphics[width=0.45\textwidth]{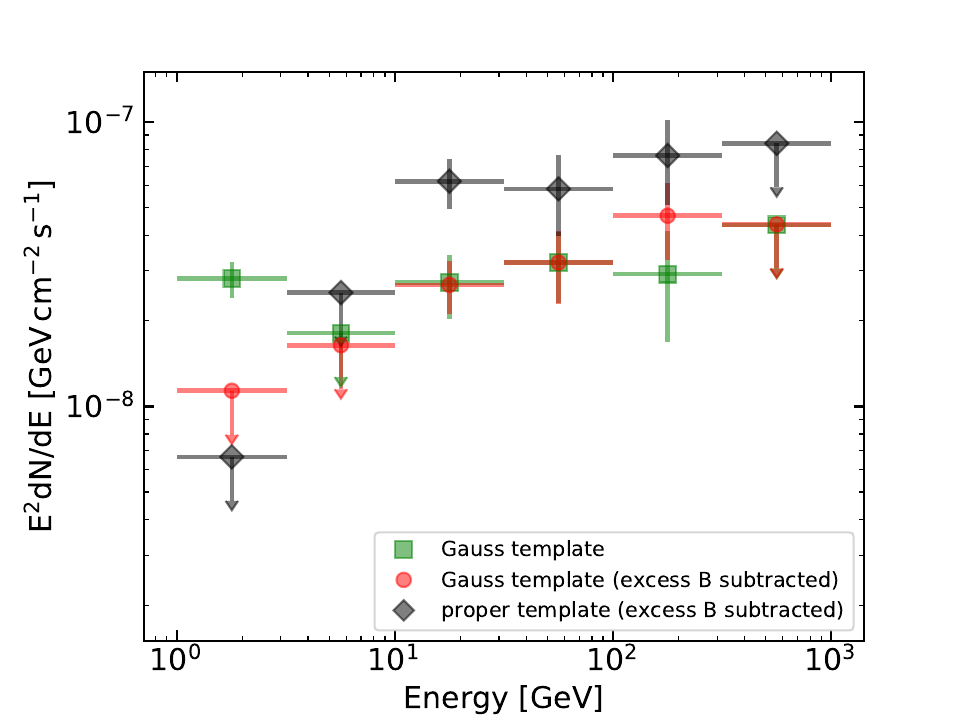}
\caption{SEDs of the Geminga halo (or say exess A) for different analyes. The Gaussian template is centered on (78$^\circ$, 7$^\circ$) with $\sigma=10^\circ$.
The proper template refers to the halo template used to generate the results shown in Fig.~\ref{fig:sed} and Table~\ref{tab:likers}.}
\hfill
\label{fig:sed3}
\end{figure}

\begin{figure*}
\centering
\includegraphics[width=0.45\textwidth]{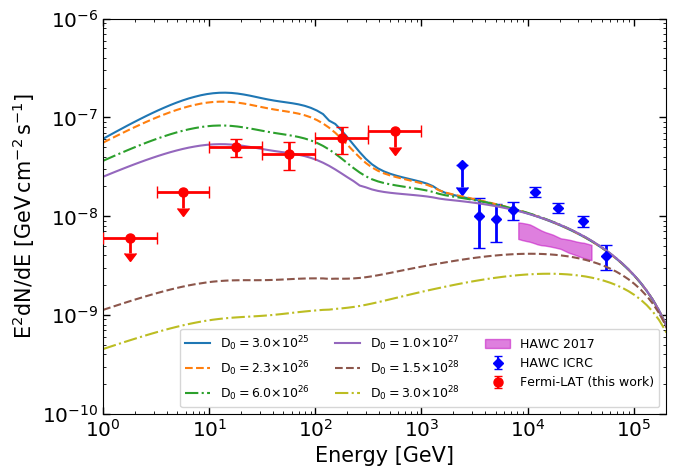}
\includegraphics[width=0.45\textwidth]{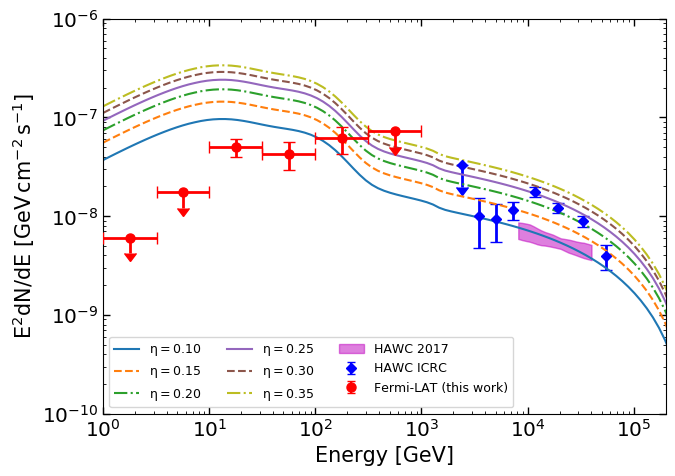}\\
\includegraphics[width=0.45\textwidth]{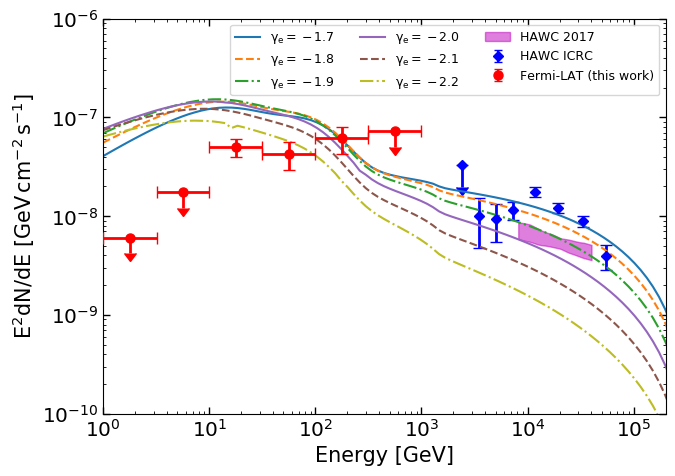}
\includegraphics[width=0.45\textwidth]{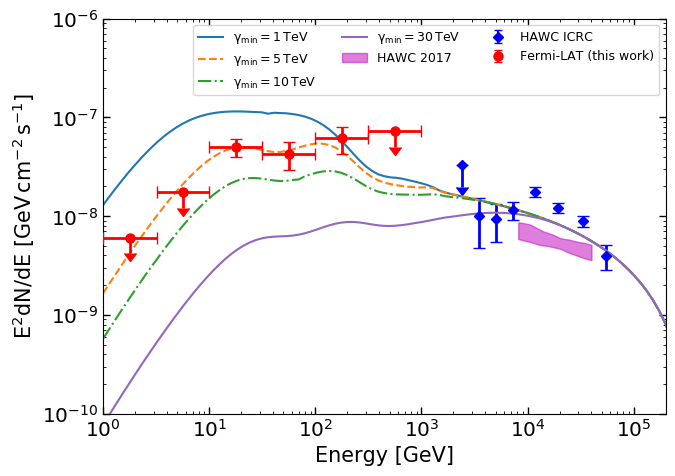}\\
\caption{Comparisons between observational data and theoretical expectation for the gamma-ray spectrum of Geminga halo. The model curves are derived by changing only the parameters of $D_0$ (upper-left panel), $\eta$ (upper-right panel) and $\gamma_{\rm min}$ (bottom-left panel) on the basis of the benchmark parameters list in Table \ref{tab:pars}. In the bottom-right panel, we additionally impose a low-energy cutoff $\gamma_{\rm min}$ of the injected $e^\pm$ spectrum on the benchmark model.}
\hfill
\label{fig:cpr}
\end{figure*}

We note that the center of this excess emission (cross point) is slightly offset from the position of Geminga's born. However, drawing a contour of 95\% C.L. (i.e., $\Delta$TS = 6.1 for 2 degrees of freedom) around the maximum position reveals that the Geminga born position is still within the 95\% localization error region (though close to the boundary).
Therefore, the TS map results are still consistent with the hypothesis that the excess emission comes from the Geminga ICS halo.
Note also that our calculation here of the Geminga born position uses a typical value of the pulsar's age as reported in the literature. By assuming that the age of Geminga is older than 342 kyr, the pulsar's born position can be adjusted to more closely match the result of the Fermi-LAT analysis. Therefore, if in the future the excess emission is confirmed to come from the ICS halo (for instance, with the observations from future larger telescopes \citep{2022AcASn..63...27F}), then the halo's observations at GeV energies can provide information on the age and braking index of the Geminga pulsar.

Assuming that the position of Geminga born is near R.A.=78$^\circ$, Dec.=7$^\circ$ (namely the position with the maximum TS value in the TS map, see Fig.~\ref{fig:tsmap1}), we alternatively derive the SED of the halo by placing the halo component (we use a $\sigma=10^\circ$ Gaussian template to model the halo in this test) at (78$^\circ$, 7$^\circ$), and the results are shown in Fig.~\ref{fig:sed3} (green points). In this case, the halo component has a higher flux near 1 GeV, which seems not to match the above results that the halo flux is low around this energy. 
The first flux point has a TS value of $\sim$61. To check the origin of this flux at low energy end, we generate another TS map with $\sigma=12^\circ$ (the $\sigma$ is also determined through the extension analysis of Appendix \ref{sec:ext}) in the energy range of 1-10 GeV (right panel of Fig.~\ref{fig:tsmap1}). We find that there is a large region of excess emission (or background redsidual) appearing in the upper right part of the TS map (denoted as excess B), and the 1-3.6 GeV flux point we see in the SED of Fig.~\ref{fig:sed3} is likely contaminated by this excess. Performing the same extension analysis and localization analysis as above, we find that the central position (i.e., the maximum position of the TS map) of excess B is about (75$^\circ$, 26$^\circ$), with an extension of about 12$^\circ$. Deriving the SED of the halo component again by additionally adding the excess B into the model (the excess B is modeled as a 12$^\circ$ Gaussian centered on (75$^\circ$, 26$^\circ$) and is simultaneously fitted together with the halo component), we obtain the spectra shown in Fig~\ref{fig:sed3} (red and black points). As is shown, after subtracting the excess B, the first point of the SED decreases to $\lesssim10^{-8}\,{\rm GeV\,cm^{-2}s^{-1}}$, consistent with the result in Fig.~\ref{fig:sed}. Therefore, even though assuming the central position of the halo component is near (78$^\circ$, 7$^\circ$), our main conclusion still holds: the Gminga ICS halo has a low flux near the GeV energies.

At last, we discuss what model parameters can match the observed SED under the assumption that the excess is true from the ICS halo. We derive model-expected spectra for different parameters of $\gamma_e$, $\eta$, and $D_0$.
Fig.~\ref{fig:cpr} shows the comparisons between the model-expected spectra and the measurements. {In the upper-left panel of Fig.~\ref{fig:cpr}, we notice that when the diffusion coefficient $D_0$ is larger than $\sim10^{28}\,{\rm cm^2/s}$, the predicted gamma-ray spectra of the halo show a relatively large difference compared to cases of $D_0 < 10^{27}\,{\rm cm^2/s}$. 
This is because diffusion becomes dominant over energy loss when $D_0\gtrsim10^{28}\,{\rm cm^2/s}$, so particles escape the halo region before producing much radiation.}
For the first three panels, one can see that the model curves always exceed the flux upper limit at energies around 1 GeV by $\sim1$ order of magnitude, and the observation can not be explained by only changing the parameters of $\gamma_e$, $\eta$, and $D_0$. 
This may also be another reason why the 1-1000 GeV analyses of Configs.~I to III in Table~\ref{tab:likers} give lower TS values, because the spectrum in the model template does not fully match the observations.
Particularly at low energies ($<$10~GeV), the model-predicted ICS spectrum fails to reproduce the observed low flux.

To fit the measured SED, we tentatively impose a low energy cutoff $\gamma_{\rm min}$ on the injected $e^\pm$ spectrum, and the resulting ICS spectrum is shown in the bottom-right panel of Fig.~\ref{fig:cpr}. 
It can be seen that if the steep rise in $E^2dN/dE$ spectrum below 10 GeV is due to a low-energy cutoff of the $e^\pm$ spectrum, a cutoff energy of $\gamma_{\rm min}>5\,{\rm TeV}$ is required to fit the observation.
If the steep rise in the spectrum is real, we currently do not know exactly what mechanism creates such a low-energy cutoff $\gamma_{\rm min}$ in the electron spectrum. It might be related to energy-dependent confinement, where low-energy electrons are trapped in the PWN and cannot escape into outer space. We note that some studies have also suggested adding a low-energy cutoff to the electron injection spectrum to address the problem of an overly large energy budget \cite{Breuhaus:2021vkr,Zhou:2022jzg}.
Please also note that we still need to further confirm that the observed halo-like excess is truly from the Geminga halo. If no reasonable physical explanation exists, the low flux at $<10\,{\rm GeV}$ may actually challenge the authenticity of the detected GeV halo.

\section{Summary}
In this work, we revisit the analysis on the ICS gamma-ray halo of the Geminga pulsar with the Fermi-LAT data. We use a longer data set (13 years) in a wider energy range (1-1000 GeV) to update previous analyses. We confirm that an extended emission component can be identified around Geminga in the energy range of 10-1000 GeV, with a maximum TS value of about $\sim$63. However, we find that 
the analysis in the 1-1000 GeV range results in a much lower TS value of the halo. 
The lower significance in the 1-1000 GeV analysis may result from the IEM model being unable to perfectly fit the background over this broader energy range, or from the template generated with the benchmark parameters not being consistent with the observations at low energies.
By derive the SED, we find that the halo component has a very low flux below 10 GeV.
Imposing a low-energy cutoff of the injected $e^\pm$ spectrum can better interpret the measured gamma-ray spectrum in the scenario of the Geminga halo.
If the detected extended excess component is real from the pulsar halo, the low-energy spectral break might reflect the energy-dependent confinement of particles from the pulsar wind nebula.

\section*{Acknowledgments}
We would like to thank Shaoqiang Xi, Ruoyu Liu, Kun Fang and Qizuo Wu for helpful discussions and suggestions.
This work has made use of data and software provided by the Fermi Science Support Center. This work is supported by the National Key Research and Development Program of China (Grant Nos. 2022YFF0503304, 2024YFA1611700), the National Natural Science Foundation of China (Grant Nos. 12203015, 12133003), and the Guangxi Talent Program (``Highland of Innovation Talents'').
	
\bibliographystyle{apsrev4-1-lyf}
\bibliography{mainNotes}

\widetext
\newpage

\appendix

\section{Phase-folded light curve of Geminga}
\label{sec:phase}
\begin{figure}[h]
\centering
\includegraphics[width=0.5\textwidth]{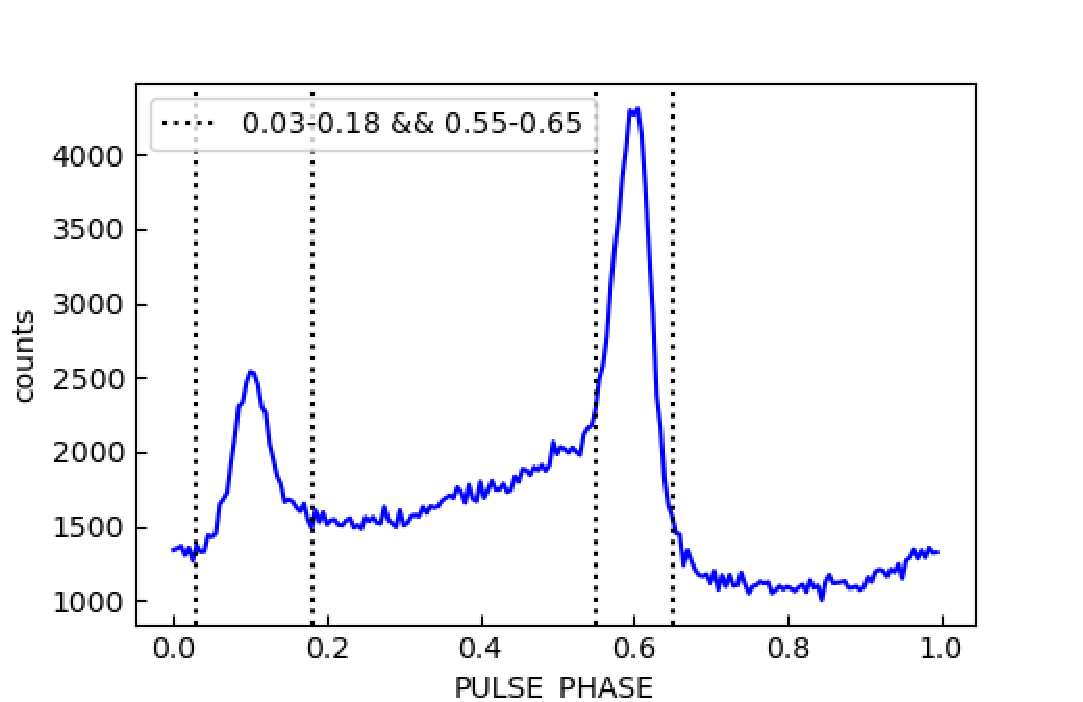}
\caption{Phase-folded light curve of Geminga in 1-5 GeV energy range. The vertical dotted lines separate the on-pulse (0.03-0.18,0.55-0.65) and the off-pulse (0-0.03,0.18-0.55,0.65-1.0) period defined in our analysis.}
\label{fig:lc}
\end{figure}

We use the TEMPO2 software \citep{tempo2} and the ephemerides of Geminga \citep{2013ApJS..208...17A} to assign a pulse phase to each photon within the ROI. 
Then we obtain a phase-folded light curve of Geminga. The 1-5 GeV light curve is shown in Figure~\ref{fig:lc}. We define $\phi\in[0.03,0.18]\cup[0.55,0.65]$ as the on-pulse period and the rest as the off-pulse period.

\begin{figure}[b]
\centering
\includegraphics[width=0.45\textwidth]{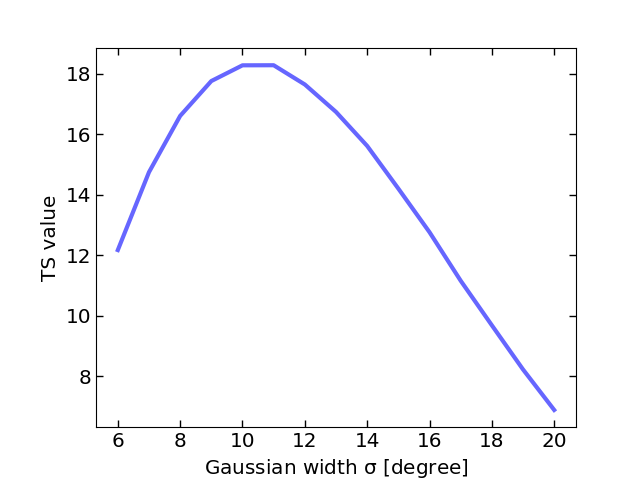}
\includegraphics[width=0.45\textwidth]{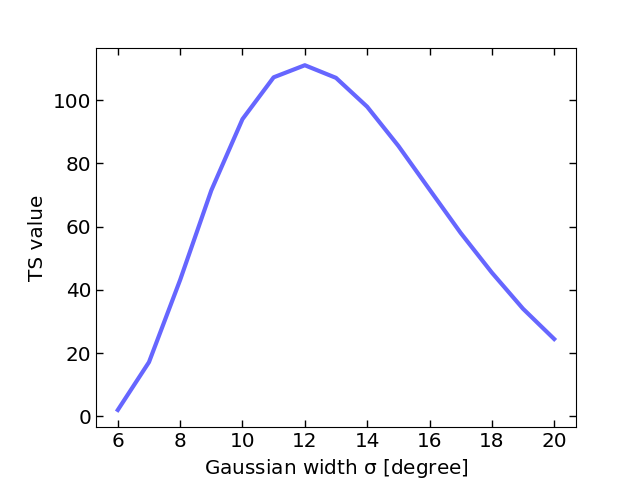}
\caption{TS value as a function of Gaussian extension $\sigma$ for the excess A in the 10-1000 GeV energy range ({\it left}) and for the excess B in the 1-10 GeV ({\it right}).}
\hfill
\label{fig:ext}
\end{figure}

\section{Extension analysis of the excess emission}
\label{sec:ext}
{The center position and extension of the Gaussian templates for modeling the excess A and B are determined by the following steps. We first use a $\sigma=15^\circ$ (this value is arbitrarily chosen before the analysis) Gaussian to generate a TS map, and find the maximum TS value of the TS map. Then we perform an extension analysis at the maximum position (i.e., plotting the curve of TS value as a function of Gaussian radius). We find that the Gaussian with $\sigma = 10^\circ/12^\circ$ gives the largest TS value for excess A and B, respectively (see Figure~\ref{fig:ext}). We use the Gaussian model with $\sigma = 10^\circ/12^\circ$ to again generate the TS map, and find that the maximum position of the obtained TS map at this time coincides with that of the $\sigma=15^\circ$ TS map, indicating that $\sigma = 10^\circ/12^\circ$ is a suitable choice. The center position of the Gaussian template used to derive the spectra of Figure~\ref{fig:sed3} is set to the maximum position of the TS map in the second step (i.e., the TS map in Figure~\ref{fig:tsmap1}).}

\section{Gamma-ray emission from the Geminga halo}
\label{sec:cal}
Here we briefly describe the formalism used to calculate the gamma-ray spectrum and morphology of the Geminga halo.
Our benchmark parameters for the calculation of the ICS halo emission and for generating the template used in the later Fermi-LAT data analysis are listed in Table \ref{tab:pars}.
\subsection{Electron diffusion}
The diffusion of $e^+/e^-$ in the space is described by the diffusion equation
\begin{align}
\frac\partial{\partial t} {N(E_e,\vec{r},t)}&=\frac{D(E_e)}{r^2}\frac{\partial}{\partial r}r^2\frac{\partial}{\partial r}N(E_e,\vec{r},t) \nonumber\\
+&\frac{\partial}{\partial E_e}\left[b(E_e)N(E_e,\vec{r},t)\right]+Q(E_e,\vec{r},t)
\label{eq1}
\end{align}
where $N$ is the differential number density of the electron. For the diffusion coefficient, we consider $D(E)=D_0(E/{\rm GeV})^\delta$ with $\delta=1/3$. 
The energy loss rate of $e^+/e^-$ is quantified by $b(E)$. At the relevant energies of this work, the electrons mainly loss energy by the synchrotron and inverse Compton scattering, and the loss rate can be approximated by $b(E)\approx b_2E^2$ with $b_2=0.6\times10^{-16}\,{\rm GeV/s}$ \citep{Atoyan1995}.
It should be noted that the energy losses have a complicated energy dependence across the GeV up to TeV energies. Moreover, the exact form of $b(E)$ in the vicinity of Geminga is not well known. These may introduce significant uncertainty in this term.

\subsection{Source injection}
The $Q(E_e,\vec{r},t)=Q(E_e, t)\delta(\vec{r}-\vec{r}_s)$ in Eq. (\ref{eq1}) is the $e^+/e^-$ injection of the pulsar,
\begin{align}
Q(E_e, t)&=L(t)\times S(E_e) \nonumber\\
&={L_{0}}{\left(1+\frac{t}{\tau_{0}}\right)^{-2}}\times S_0\left(\frac{E_e}{E_{0}}\right)^{-\gamma_e} \exp \left(-\frac{E_e}{E_{\rm max}}\right).
\end{align}
The time evolution related term $L(t)$ is due to the magnetic dipole braking of the pulsar, and we assume that the magnetic braking index is $k=3$ and the spin-down time scale is $\tau_0=12\,{\rm kyr}$. The initial injection power is related to the total energy released by the pulsar $L_0 = \eta E_{\rm tot}/\tau_0$, with $\eta$ the conversion efficiency. We adopt $E_{\rm tot}=1.05\times10^{49}\,{\rm erg}$ in our calculation.
The $S_0$ is introduced to normalize the spectrum-related term, $\int E_e S(E_e){\rm d}E_e=1$.

\subsection{Solution to the diffusion equation}
The diffusion equation (Eq. (1)) can be solved with the Green's function method,
\begin{align}
N\left(E_{e}, \vec{r}, t\right)=&\int_0^{t} d t_{0} \frac{b\left(E_{e}^*\right)}{b\left(E_{e}\right)} \frac{1}{\left[\pi \lambda^{2}\left(t_{0}, t, E_e\right)\right]^{\frac{3}{2}}} \nonumber\\
&\times\exp \left(-\frac{\left|\vec{r}-\vec{r}_s\right|^{2}}{\lambda^{2}\left(t_{0}, t, E_e\right)}\right) Q\left(E_{e}^*, t_{0}\right)
\end{align}
where $\vec{r}_s$ is the position of the source. In the case of taking into account pulsar's proper motion, the $\vec{r}_s$ is time dependent, $\vec{r}_s(t)=\vec{r}_0+\vec{v}t$, where $\vec{v}$ is the pulsar velocity and $\vec{r}_0$ is the initial position when the pulsar is born.
For the Geminga pulsar, the proper motion is 178.2$\pm$1.8 mas/year \citep{ATNFpaper,Caraveo96_distpm}.
The $E_e^*$ is the initial energy of electron that cool down to $E_e$ in a loss time of $\Delta\tau=t-t_0$:
\begin{equation}
E_e^*\simeq \frac{E_{e}}{\left[1-b_{2} E_{e}\left(t-t_{0}\right)\right]}.
\end{equation}
And the diffusion length is given by
\begin{equation}
\lambda(E)\approx 2\sqrt{\frac{D_0[E_e^{*(\delta-1)}-E_e^{(\delta-1)}]}{ b_2(\delta-1)}}.
\end{equation}

\subsection{Seed photon field of ICS}
The halo gamma rays are produced through inverse Compton scattering off CMB, infrared (IR), and optical photons. 
The photon fields are considered to have a graybody distribution of which the temperature and energy density are approximately those derived by GALPROP \citep{Moskalenko98galprop}. 
For the position of Geminga, the temperatures and energy densities of the CMB, IR and optical components are (2.7 K, $0.26\,{\rm eV/cm^3}$), (20 K, $0.3\,{\rm eV/cm^3}$) and (5000 K, $0.3\,{\rm eV/cm^3}$), respectively \citep{hawc17}. The ICS is computed using the GAMERA package\footnote{\url{http://libgamera.github.io/GAMERA/}} and the standard formula given in \cite{Blumenthal70} is adopted.

\begin{table}[h]
\caption{Benchmark parameters for the halo ICS emission.}
\begin{ruledtabular}
\begin{tabular}{ccl}
  Parameters & Value & Description \\
\hline
$t_a$        &  342 kyr  & Pulsar age \\
$d$   &  250 pc  & Distance \\
$v$   &  211 km/s  & Velocity of proper motion\\
$\dot{E}$  &  $3.2\times 10^{34}\,{\rm erg/s}$   &  Spin-down power \\
$\tau_0$ & 12 kyr  & Spin-down timescale \\
$k$  & 3 & Break index \\
$\eta$ &  0.15  &  Conversion efficiency  \\
\hline
$D_0$ &  $2.3\times10^{26}\,{\rm cm^2/s}$  & $D(E)$ at 1 GeV\\
$\delta$ &  1/3  & Slope of $D(E)$ \\
\hline
$\gamma_e$  & -1.8 & $e^\pm$ spectral index \\
$E_{\rm max}$  & 1000 TeV & Cutoff energy of $e^\pm$\\
\end{tabular}
\end{ruledtabular}
\label{tab:pars}
\end{table}

\end{document}